\documentclass[twocolumn,prl,amsmath,superscriptaddress,nofootinbib,amssymb]{revtex4}
\usepackage{graphicx}
\usepackage{dcolumn}
\usepackage{bm}
\usepackage[usenames]{color}
\usepackage[normalem]{ulem}
\usepackage{float}
\usepackage{epstopdf}
\usepackage{soul}

\begin{document}

\title{Termination-dependent surface properties\\
 in the giant-Rashba semiconductors BiTe$X$ ($X$\,=\,Cl, Br, I)}

\author{Sebastian Fiedler}
\affiliation{Experimentelle Physik VII and R\"ontgen Research Center for Complex Materials (RCCM), Universit\"at W\"urzburg, Am Hubland, D-97074 W\"urzburg, Germany}

\author{Thomas Bathon}
\affiliation{Experimentelle Physik II and R\"ontgen Research Center for Complex Materials (RCCM), Universit\"at W\"urzburg, Am Hubland, D-97074 W\"urzburg, Germany}

\author{Sergey V. Eremeev}
\affiliation{Institute of Strength Physics and Materials Science, 634055, Tomsk, Russia}
\affiliation{Tomsk State University, 634050, Tomsk, Russia}
\affiliation{Saint Petersburg State University, 198504, Saint Petersburg, Russia}

\author{Oleg E. Tereshchenko}
\affiliation{Saint Petersburg State University, 198504, Saint Petersburg, Russia}
\affiliation{Institute of Semiconductor Physics, 636090, Novosibirsk, Russia}
\affiliation{Novosibirsk State University, 636090, Novosibirsk, Russia}

\author{Konstantin A. Kokh}
\affiliation{Saint Petersburg State University, 198504, Saint Petersburg, Russia}
\affiliation{Novosibirsk State University, 636090, Novosibirsk, Russia}
\affiliation{Institute of Geology and Mineralogy, SB RAS, 630090, Novosibirsk, Russia}

\author{Evgueni V. Chulkov}
\affiliation{Tomsk State University, 634050, Tomsk, Russia}
\affiliation{Saint Petersburg State University, 198504, Saint Petersburg, Russia}
\affiliation{Donostia International Physics Center (DIPC), 20018 San Sebasti\'an/Donostia, Basque Country, Spain}
\affiliation{Departamento de F\'isica de Materiales and Centro Mixto CSIC-UPV/EHU, Facultad de Ciencias Qu\'{\i}micas, Universidad del Pais Vasco/Euskal Herriko Unibertsitatea, Apdo. 1072, 20080 San Sebasti\'an/Donostia, Basque Country, Spain}

\author{Paolo Sessi}
\affiliation{Experimentelle Physik II and R\"ontgen Research Center for Complex Materials (RCCM), Universit\"at W\"urzburg, Am Hubland, D-97074 W\"urzburg, Germany}

\author{Hendrik Bentmann}
\affiliation{Experimentelle Physik VII and R\"ontgen Research Center for Complex Materials (RCCM), Universit\"at W\"urzburg, Am Hubland, D-97074 W\"urzburg, Germany}

\author{Matthias Bode}
\affiliation{Experimentelle Physik II and R\"ontgen Research Center for Complex Materials (RCCM), Universit\"at W\"urzburg, Am Hubland, D-97074 W\"urzburg, Germany}

\author{Friedrich Reinert}
\affiliation{Experimentelle Physik VII and R\"ontgen Research Center for Complex Materials (RCCM), Universit\"at W\"urzburg, Am Hubland, D-97074 W\"urzburg, Germany}

\date{\today}

\begin{abstract}
The non-centrosymmetric semiconductors BiTe$X$ ($X$ = Cl, Br, I)
show large Rashba-type spin-orbit splittings in their electronic
structure making them candidate materials for spin-based electronics.
However, BiTeI(0001)
single crystal surfaces usually consist of stacking-fault-induced
domains of Te and I terminations implying a spatially inhomogeneous
electronic structure. Here we combine scanning tunneling microscopy (STM),
photoelectron spectroscopy (ARPES, XPS) and density functional theory (DFT)
calculations to systematically investigate the structural and
electronic properties of BiTe$X$(0001) surfaces. For $X$ = Cl, Br we
observe macroscopic single-terminated surfaces. We discuss chemical
characteristics among the three materials in terms of bonding character,
surface electronic structure, and surface morphology.
\end{abstract}

\maketitle

\section{Introduction}

The narrow-gap semiconductors BiTe$X$ ($X$ =  Cl, Br, I) have attracted considerable
interest because of large Rashba-type spin-orbit splittings in their bulk and surface electronic structures \cite{Ishizaka, Eremeev_PRL, Murakawa}, which have been observed by angle-resolved photoelectron spectroscopy (ARPES) \cite{Crepaldi_PRL, Landolt_PRL,Landolt_NJP, Sakano_PRL, Moreschini} and magnetotransport measurements \cite{Martin, Bell}.
The enhanced spin-splitting in these materials is driven by their non-centrosymmetric crystal structure in combination with strong atomic spin-orbit coupling and a negative crystal-field splitting of the topmost valence bands \cite{Bahramy_PRB}.
The latter features have also been predicted to promote a topological insulator phase in BiTeI under application of external pressure \cite{Bahramy}. The BiTe$X$ series does not only host the presently largest known Rashba effect of all semiconductors, it also appears more suitable for possible spin-electronic applications \cite{Datta, Zutic} than artificially grown monolayer reconstructions, such as metallic surface alloys, where spin-splittings of similar magnitude can be achieved \cite{Ast, Bentmann, El-Kareh_PRL, El-Kareh_NJP}.

At the surface, the non-centrosymmetric, layered unit cell of BiTe$X$ results in two possible polar terminations \cite{Crepaldi_PRL,Eremeev_NJP,Landolt_NJP, Moreschini}, Te- and $X$-terminated surfaces, that give rise to n-type or p-type band bending, respectively \cite{Crepaldi_PRL}. The surface properties may be influenced additionally by defects, as is the case for BiTeI, where bulk stacking faults induce coexisting Te- and I-terminated domains on microscopic length scales as shown by scanning tunneling microcopy (STM) \cite{Butler, Tournier, Kohsaka, Fiedler}.
While the resulting lateral interfaces between surface areas of different terminations may provide interesting new physics \cite{Butler, Tournier}, the presence of multiple domains will in most instances be undesirable. For BiTeCl and BiTeBr spatially resolved surface investigations have so far been scarce \cite{Yan}. In the case of BiTeCl photoemission experiments indicate single-terminated surfaces \cite{Landolt_NJP}, in contrast to BiTeI, whereas for BiTeBr the situation is unclear. The majority of ARPES studies of BiTe$X$ point to similar Rashba-split band structures for all three compounds \cite{Sakano_PRL, Landolt_NJP, Crepaldi_PRB, Moreschini}, in agreement with theoretical predictions \cite{Eremeev_PRL, Moreschini}. However, for BiTeCl the existence of topological surface states has also been claimed based on ARPES \cite{Chen} and STM \cite{Yan}.

In this work we present a combined investigation of the surface
structural and electronic properties of the BiTe$X$ semiconductors.
Our STM experiments show that BiTeBr and BiTeCl(0001) display
single-domain surfaces with X- or Te-termination. The determined
terrace step heights agree with the respective bulk unit cell
parameters and X-ray
photoemission (XPS) provides depth-dependent chemical information in
line with the expected layered atomic structure. The measured
core-level binding energies indicate a significant charge transfer
from Bi to, both, $X$ and Te in agreement with density functional
theory (DFT) calculations. We systematically compare the electronic
properties of Te- and $X$-terminated surfaces in terms of band
bending, surface band structure, work function, atomic defects, and
reaction to deposited adsorbate atoms.

\section{Methods}

Our experimental setup is designed for a comprehensive analysis of
the geometric and electronic properties in real and reciprocal
space as described in Ref.\,\cite{Fiedler}. The system allows
surface analytics by means of various techniques, i.e.
LEED, SPA-LEED, STM, STS, AFM, XPS, work function and ARPES
measurements in
ultra-high vacuum conditions for the same sample without exposing it
to air. Additional high-resolution STM measurements were performed
at a separate setup with a low-temperature STM (Omicron LT-STM) at
$T=5$\,K.

We used a modified sample holder system, which allows to split single crystals
\textit{in situ} and to measure both corresponding surfaces of a cleave
without the need to re-glue or to expose the sample to air [see Fig.~\ref{Figure1}(a)]. Thus, BiTe$X$ ($X$ = Cl, Br, I) single crystals were cleaved at room temperature along the (0001) direction at pressures low $2\cdot10^{-10}$\,mbar revealing surfaces of about 2\,mm$\times$2\,mm
on each side. A podium smaller than the sample was used to
move the surface into the focal point of the electron spectrometer in order to minimize spurious signal from the sample holder.

Submonolayer amounts of Cs
were deposited using commercial alkali dispensers (SAES Getters).
All experiments were performed at room temperature except for those
carried out at the LT-STM.

Tips have been prepared according to Ref.\,\cite{Fiedler}. Differential conductance maps are used to obtain spatially resolved information about the sample's local density of states (DOS).  For this purpose a small modulation voltage ($U_{\rm mod} = 25$\,mV) is added to the sample bias $V$ and the resulting variation of the tunneling current, $\mathrm{d}I/\mathrm{d}V$, is recorded simultaneously with the topograhic, i.e.\ constant-current image. STM data were processed with the WSxM software package \cite{Horcas}.

XPS measurements were done with Al K$\alpha$ radiation ($h\nu$~=~1486.6\,eV) under a photoelectron emission angle of 60\,$^\circ$ in order to enhance the surface sensitivity of the experiment. The X-ray source was not monochromatized and the spectra were satellite-corrected. The energy resolution of the XPS measurements was ca.~1~eV. ARPES data were acquired with a non-monochromatized He discharge lamp with He I$\alpha$ radiation ($h\nu$~=~21.2\,eV) and at an energy resolution of approximately 25\,meV. Work functions were determined from the secondary photoelectron cutoff with the sample held on a positive potential of 9\,V. Calibration measurements for Au(111) gave values in line with previous reports \cite{Trasatti, Rusu}.

The synthesis of the charges was performed by fusing binary compounds: Bi$_2$Te$_3$ with BiCl$_3$, BiBr$_3$ and BiI$_3$, correspondingly. According to published data \cite{Tomokiyo, Petasch} BiTeI and BiTeBr melt congruently at 560\,$^\circ$C and 526\,$^\circ$C, while BiTeCl shows incongruent melting \cite{Petasch} at 430\,$^\circ$C with a peritectic composition around 11\,mol.\% Bi$_2$Te$_3$ +\,89 mol.\% BiCl$_3$. Therefore we have used stoichiometric charge for BiTeI, BiTeBr and melt-solution system with a molar ratio Bi$_2$Te$_3$:BiCl$_3$\,=\,1:9 for the crystallization of BiTeCl. The synthesis was performed directly in the growth quartz ampoules at a temperature which is 20\,$^\circ$C above the melting point. Crystal growth was performed by the modified Bridgman method with rotating heat field \cite{CrystEngComm}. After pulling the ampoules through the vertical temperature gradient of 15\,$^\circ$C/cm at 10\,mm/day, the furnace was switched off.

Complementary first-principles calculations were performed within
the framework of the density functional theory (DFT) using the
projector-augmented-wave (PAW) \cite{PAW1,PAW2} basis. The
generalized gradient approximation (GGA-PBE) \cite{PBE} to the
exchange correlation (XC) potential as implemented in the {\sc VASP}
code \cite{VASP1,VASP2} was used. The relaxed bulk parameters have
been taken into account. The atomic charges
were estimated by implementing the Bader charge
analysis \cite{Bader}.

\section{Results}

\subsection{Surface morphology and bonding character}

%==================================================================
\begin{figure*}
\begin{center}
\includegraphics[width=0.8\textwidth]{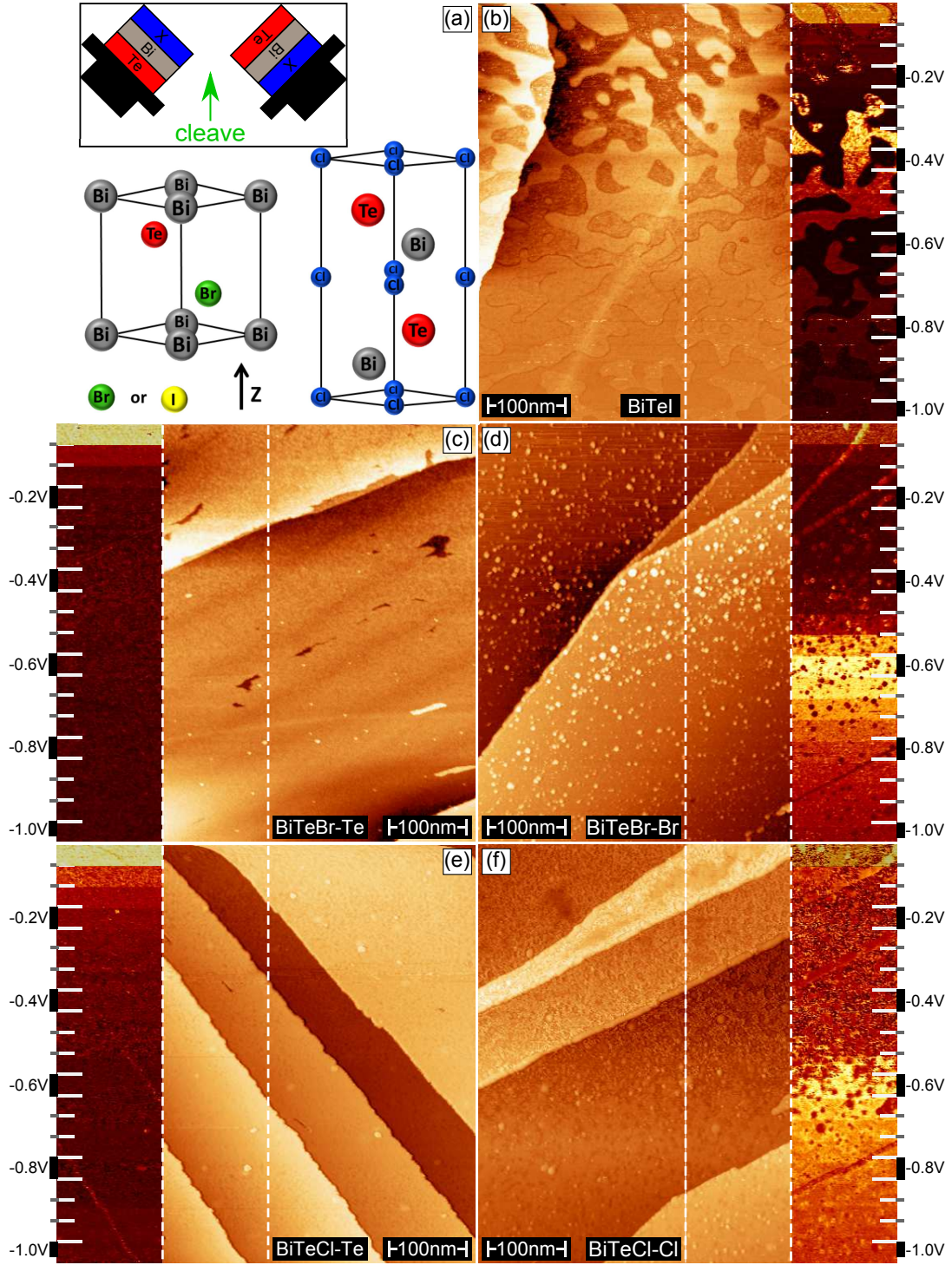}
\end{center}
\caption{
Crystal structure and room-temperature STM measurements for BiTe$X$. (a) Bulk unit cells of  BiTeBr/I and BiTeCl and the resulting surface terminations after cleaving. The inset sketches the situation for an ideal crystal mounted between the two sample holders (black) after the cleave. STM measurements (500\,nm$\times$500\,nm) of (b) BiTeI, (c) Te-terminated and (d) Br-terminated BiTeBr, as well as (e) Te-terminated and (f) Cl-terminated BiTeCl. The gap voltage is varied from -0.05\,V to -1\,V and the tunneling current was 0.1\,nA in (b) and 0.2\,nA in (c)-(f). The outmost parts of the images are d\textit{I}/d\textit{V} maps of the areas between the lines.}
\label{Figure1}
\end{figure*}
%==================================================================

Fig.~\ref{Figure1}(a) shows the unit cells of BiTe$X$. While BiTeI
and BiTeBr have a unit cell of 3 atomic layers, the one of BiTeCl comprises 6 layers along $z$
resulting in a height twice as large \cite{Eremeev_PRL,Shevelkov}.
The inset sketches the stacking order after the cleave of an ideal single crystal,
resulting in two different terminations for the two opposing surfaces.

Fig.~\ref{Figure1}(b) displays a 500\,nm$\times$500\,nm STM
measurement of BiTeI(0001) at 0.1\,nA tunneling current.
During the scan the gap voltage was gradually decreased from -0.05\,V
at the upper part of the image down to -1.0\,V at the lower part.
Note that negative voltages refer to tunneling from the sample to the tip,
thus reflecting the occupied DOS of the sample as being also accessed by ARPES spectra.
Coexisting Te- and I-terminations are visible as reported earlier \cite{Butler, Tournier, Kohsaka, Fiedler}.
The outer part of the image shows the corresponding d\textit{I}/d\textit{V} map of the surface
within the two white dashed lines. The Te-terminated surface shows a high
DOS at -0.05\,V while at -0.3\,V the same
surface appears dark in the d\textit{I}/d\textit{V} map and the I-terminated
surface reveals a high intensity. This high DOS originates from the
onsets of the band structures of the two different terminations, as shown in Ref. \cite{Fiedler}. The step edges within the
same terminations are around 0.7\,nm high and the ratio between Te- to I-terminated areas is
roughly 50/50.

Next we investigate the surface morphologies of BiTeBr and BiTeCl [see Figure~\ref{Figure1}(c)-(f)]. 
The images reflect a surface area of 500\,nm$\times$500\,nm, and
were obtained at 0.2\,nA tunnelling current at a voltage
varied from -0.05\,V to -1\,V. The surface terminations are indicated in the figures by
BiTeBr-Te and BiTeBr-Br for the Te- and Br-terminated surfaces of BiTeBr and by
BiTeCl-Te and BiTeCl-Cl for the Te- and Cl-terminated surfaces of BiTeCl, respectively.

Fig.~\ref{Figure1}(c) shows one side of a BiTeBr crystal split
at (0001) direction. On this surface there is no sign of a second
termination as seen in Fig.~\ref{Figure1}(b) for BiTeI. The step
edges are (0.65$\pm$0.05)\,nm high, which is in agreement with the bulk unit cell
height along $z$ \cite{Shevelkov}. Some
adsorbates can be seen but the surface is mostly clean. An increase
in the DOS close to $E_{\rm F}$ indicates that we are dealing with
the Te termination of BiTeBr, as has been shown for BiTeI in
Fig.~\ref{Figure1}(b) and Ref.\,\cite{Fiedler}. d\textit{I}/d\textit{V} maps taken over a larger energy range (not shown) further showed an onset of valence states at an energy of approximately -1\,eV.
Fig.~\ref{Figure1}(d)
shows the other side of the cleave. More adsorbates can be found on
this surface, which indicates a higher chemical reactivity. The
d\textit{I}/d\textit{V} map strongly deviates from the one obtained for the
Te-termination. At a gap voltage of around -0.55\,eV an increase in
the DOS can be seen, indicating a band onset, as observed similarly
for the I-termination of BiTeI in Fig.~\ref{Figure1}(b).
Furthermore, the adsorbates appear dark in the d\textit{I}/d\textit{V} and
start accumulating at the step edges before covering the terraces.
The higher chemical reactivity and the determined DOS indicates that
this surface is Br-terminated. For BiTeCl similar observations in
terms of DOS and adsorbate characteristics are obtained as for
BiTeBr. The STM images and d\textit{I}/d\textit{V} maps for the Te- and
Cl-terminated surface are shown Fig.~\ref{Figure1}(e) and (f),
respectively, closely resembling their counterparts in BiTeBr.
Notably, most of the step edges have a heights of (1.25$\pm$0.05)\,nm for both terminations,
matching again the height of the bulk unit cell \cite{Shevelkov}, while only 5\%--10\% of the steps
have a height of $\approx$0.7\,nm, corresponding to a single BiTeCl trilayer.

Our STM measurements thus reveal strikingly different surface
morphologies for BiTeBr and BiTeCl as compared to BiTeI. Both
compounds feature single-domain (0001) surfaces with either Te- or
$X$-termination. Apparently, bulk stacking faults, giving rise to
domains of different stacking order in BiTeI, are largely absent in
the other two compounds. A possible explanation for this behavior
could be the similar atomic radii of Te and I atoms, that might be
expected to promote the formation of mixed Te/I layers during the
crystal growth. Our DFT calculations indicate that the formation energy for stacking
faults in the bulk is much smaller for BiTeI (1\,meV) than for
BiTeBr (46\,meV) and BiTeCl (60\,meV), in line with the experimental findings.
In general, BiTeBr and BiTeCl will thus be more
suitable materials for spatially-averaging techniques that address
the spin-polarization of the electronic bulk states.

\begin{figure*}
\begin{centering}
\includegraphics[width=0.8\textwidth]{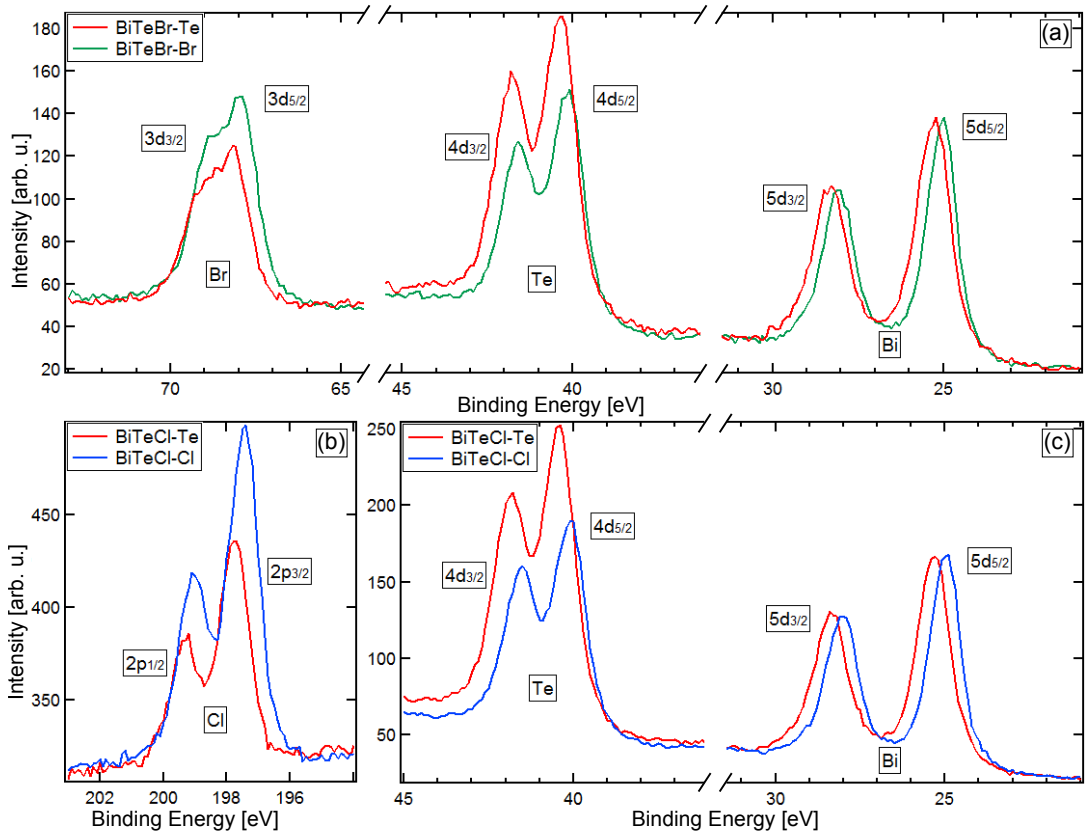}
\end{centering}
\caption{X-ray photoemission data for BiTeBr in (a) and BiTeCl in (b), (c). Characteristic intensity differences in the Te and Br/Cl core level signals are observed for the different surface terminations, reflecting the changed atomic stacking orders and the finite probing depth of the experiment. Furthermore, band bending gives rise to small energy shifts between the spectra for Te-terminated and Br/Cl-terminated surfaces.}
\label{Figure2}
\end{figure*}

To gain further insight into the structural and chemical properties of the BiTeBr and BiTeCl(0001) surfaces we have performed XPS experiments. Fig.~\ref{Figure2}(a)-(c) shows core-level spectra directly corresponding to the different surfaces presented in Fig.~\ref{Figure1}.
Comparing spectra for Te- and Br(Cl)-terminated surfaces we observe a relative shift of 200\,meV (300\,meV), which we attribute to band bending \cite{Crepaldi_PRL, Landolt_NJP, Moreschini}. The energy shifts are slightly reduced compared to values reported in Ref.\,\cite{Moreschini} which might be due to the higher excitation energy and thus an increased probing depth in the present experiments.

%==================================================================
%==================================================================

\begin{table*}
\begin{tabular}{cccccccc}
       & elem.[eV] & Bi$_2$Te$_3$[eV]   & Cl[eV] & Te$_{Cl}$[eV] & Br[eV] & Te$_{Br}$[eV] & BiTeI[eV]\\
\hline \hline
Bi 5d$_{5/2}$ &  24.1 & 24.6            &  25.0   &  25.3       &  25.0   &  25.2       &  25.0     \\
Te 4d$_{5/2}$ &  40.5 & 39.9             &  40.1   &  40.4       &  40.1   &  40.3       &  40.1     \\
\hline
work function & & 5.1     & 6.2  & 4.5   & 6.0        &4.7         & (5.2)       \\
\end{tabular}
\caption{Core level binding energies and work functions for BiTe$X$ and Bi$_2$Te$_3$. The estimated uncertainty of the measured values amounts to $\pm$0.1~eV. For comparison we also show the corresponding binding energies for elemental Bi and Te metal taken from Ref.\,\cite{Shalvoy}.}
\label{Table1}
\end{table*}

\begin{table}
\centering
\begin{tabular}{ccccc}
   & BiTeCl & BiTeBr & BiTeI \\
\hline \hline
Bi & -1.09  & -1.01  & -0.91 \\
Te & +0.41  & +0.42  & +0.44 \\
$X$& +0.68  & +0.59  & +0.47 \\
\end{tabular}
\caption{Calculated charge transfer based on DFT in the bulk BiTe$X$ compounds
(in electrons).}
\label{Table2}
\end{table}

Considering the peak intensities for the Te and Br(Cl) species we observe characteristic differences between two surfaces with different termination resulting from the finite electron mean free path of the XPS experiment of around 1\,nm \cite{Huefner}. When going from Te- to $X$-terminated surfaces the Te signal is reduced while the $X$ signal is enhanced, directly reflecting the changed atomic stacking sequence. The spectra have been normalized to the signal of Bi which for both terminations is expected to reside in the second atomic layer as shown in the inset of Fig.~\ref{Figure1}(a). For a quantitative estimation we assume an exponential damping of the signal which amounts to roughly 30\% for two atomic layers and the present experimental conditions \cite{Huefner}. From the data in Fig.~\ref{Figure2}(a) we infer that the Te 4d and Br 3d signals change by 22\% and 25\%, respectively. In Fig.~\ref{Figure2}(b)-(c) the change for the Te 4d core level is 30\% and 36\% for Cl 2p. Averaged over four samples, the damping for BiTeBr is 26$\pm$5\% for Te- and 19$\pm$6\% for Br-terminations while for BiTeCl we find 32$\pm$3\% for Te- and 24$\pm$13\% for Cl-terminated surfaces. The XPS data thus confirm the single termination and the expected termination-dependent atomic layer stacking for BiTeBr and BiTeCl.

%==================================================================
\begin{figure*}
\begin{center}
\includegraphics[width=0.8\textwidth]{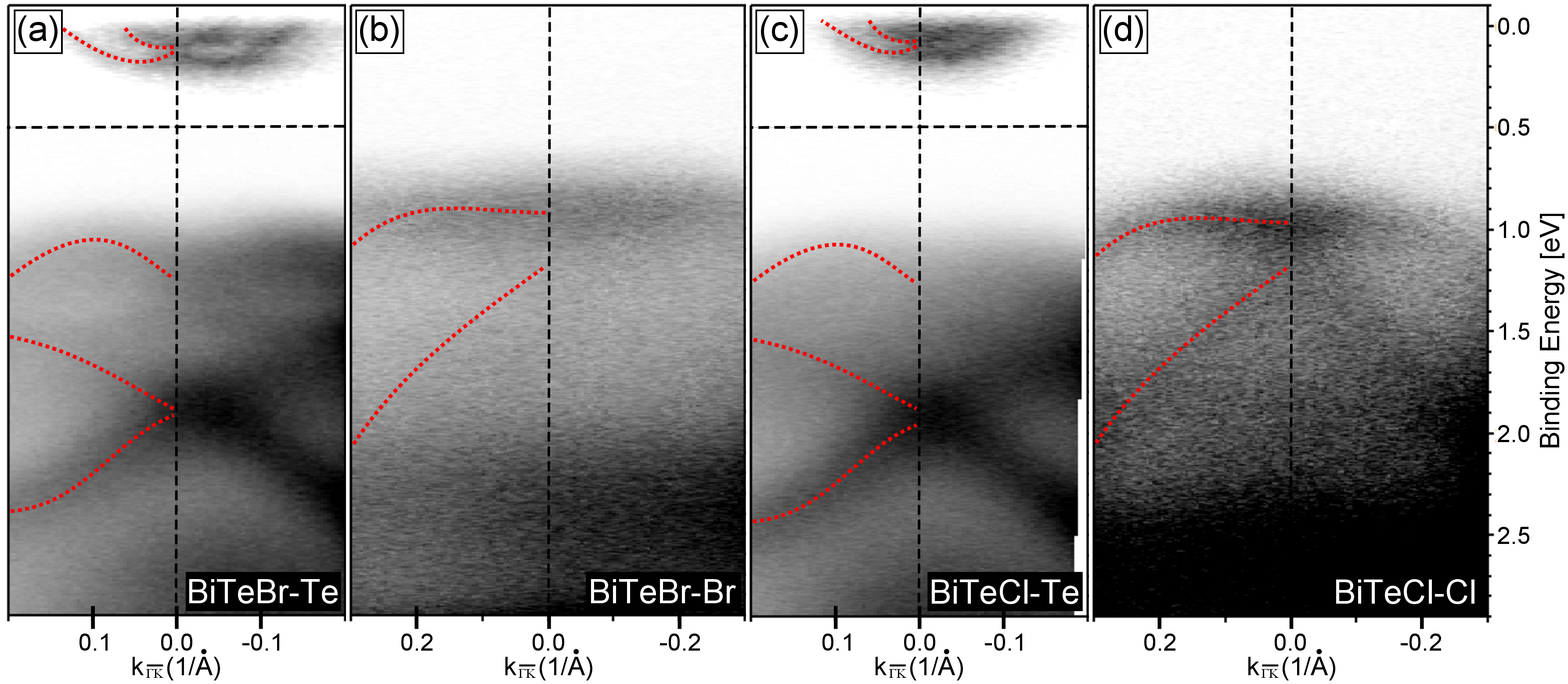}
\end{center}
\caption{Angle-resolved photoemission data for BiTeBr in (a)-(b) and for BiTeCl in (c)-(d) ($h\nu =$~21.2 eV).
The contrast between $E_{\rm F}$ and 0.5\,eV (indicated by horizontal
lines) has been increased in (a) and (c) for better visibility of
the Rashba-split band near $E_{\rm F}$.
The red dotted lines serve as guide-to-the-eye.
}
\label{Figure3}
\end{figure*}
%==================================================================

Table~\ref{Table1} summarizes the binding energies for the Te 4d$_{5/2}$
and Bi 5d$_{5/2}$ peaks in BiTe$X$, which contain information about the chemical bonding
in the compounds \cite{Huefner}. The aforementioned band bending
gives rise to small deviations between different terminations on the
order of 200-300\,meV. Furthermore, when compared to the values in
Bi and Te metal \cite{Shalvoy}, the Bi 5d$_{5/2}$ peaks are shifted
to higher and the Te 4d$_{5/2}$ peaks to lower binding energies. The
absolute shift is significantly larger for Bi than for Te.
On the other hand, no clear trends along the series $X$ = Cl,
Br, I are apparent.
To gain a better understanding of the experimental data we
have calculated by use of DFT the charge transfer in bulk BiTe$X$ as shown in
Table~\ref{Table2}. As one can see, the Bi atom loses about one electron
by transferring it to Te ($\sim$0.4\,$e$) and $X$ atoms which is in
line with the experimental result. Note that among the three
compounds the values for Bi vary by only 10 - 20\% and are basically
the same for Te. This might explain the absence of clear chemical
trends in the respective XPS binding energies. The considerable
increase in the calculated charge transfer to $X$ along $X$ = I, Br,
Cl further indicates an increasingly ionic bonding character between
$X^{-}$ and BiTe$^{+}$ layer with rising electronegativity of the
halogen atoms.

Additional insight into the influence of the
halogen species on the bonding character may be gained by a
comparison to Bi$_2$Te$_3$, showing a similarly layered structure as
BiTe$X$, where a single Bi layer resides between two Te layers (see e.g. Ref.~\cite{Kuznetsov}). For
this compound the chemical shift of the Bi 5d$_{5/2}$ line is
considerably reduced (see Table~\ref{Table1}). This points to
significant differences between BiTe$X$ and Bi$_2$Te$_3$, for which
the bonding is usually assumed to be dominated by covalent
contributions \cite{Wagner}.

Table~\ref{Table1} also displays work functions for BiTe$X$ as
determined by the secondary photoelectron cutoff. For $X$\,=\,Cl, Br
large differences above 1\,eV between $X$- and Te-terminated
surfaces are observed in quantitative agreement with a recent STM
study of the local work function on BiTeI(0001) \cite{Kohsaka}. This
finding may indeed be understood in terms of an ionic bonding
between $X^{-}$ and BiTe$^{+}$ layers creating opposite dipoles near
the surface depending on termination \cite{Shevelkov, Eremeev_NJP,
Crepaldi_PRL, Tournier, Moreschini, Butler, Ishizaka}. Furthermore,
the larger calculated charge transfer in BiTeCl compared to BiTeBr
is in line with the increased work function difference between
the two terminations observed experimentally. The work function for
a Bi$_2$Te$_3$(0001) surface, which is terminated by a Te layer \cite{Kuznetsov}, is
considerably larger than for the Te-terminated BiTeBr and BiTeCl
surfaces, again pointing to a strong effect of the halogen atoms on
the microscopic charge distribution. Surprisingly, for BiTeI only
one cutoff could be observed in our spectra despite the presence of
Te- and I-terminated surface areas. The Te and I domains of BiTeI
are in order of 100\,nm \cite{Fiedler} and maybe small enough
to result in a mixed work function when measured by
secondary electron cutoff technique. The corresponding work function
of 5.2~eV is given in brackets in Table~\ref{Table1} and lies in between
the values found for Te- and I-terminated surface areas by STM
\cite{Kohsaka}.

\subsection{Surface electronic structure}

%==================================================================

\begin{figure*}
\begin{centering}
\includegraphics[width=0.8\textwidth]{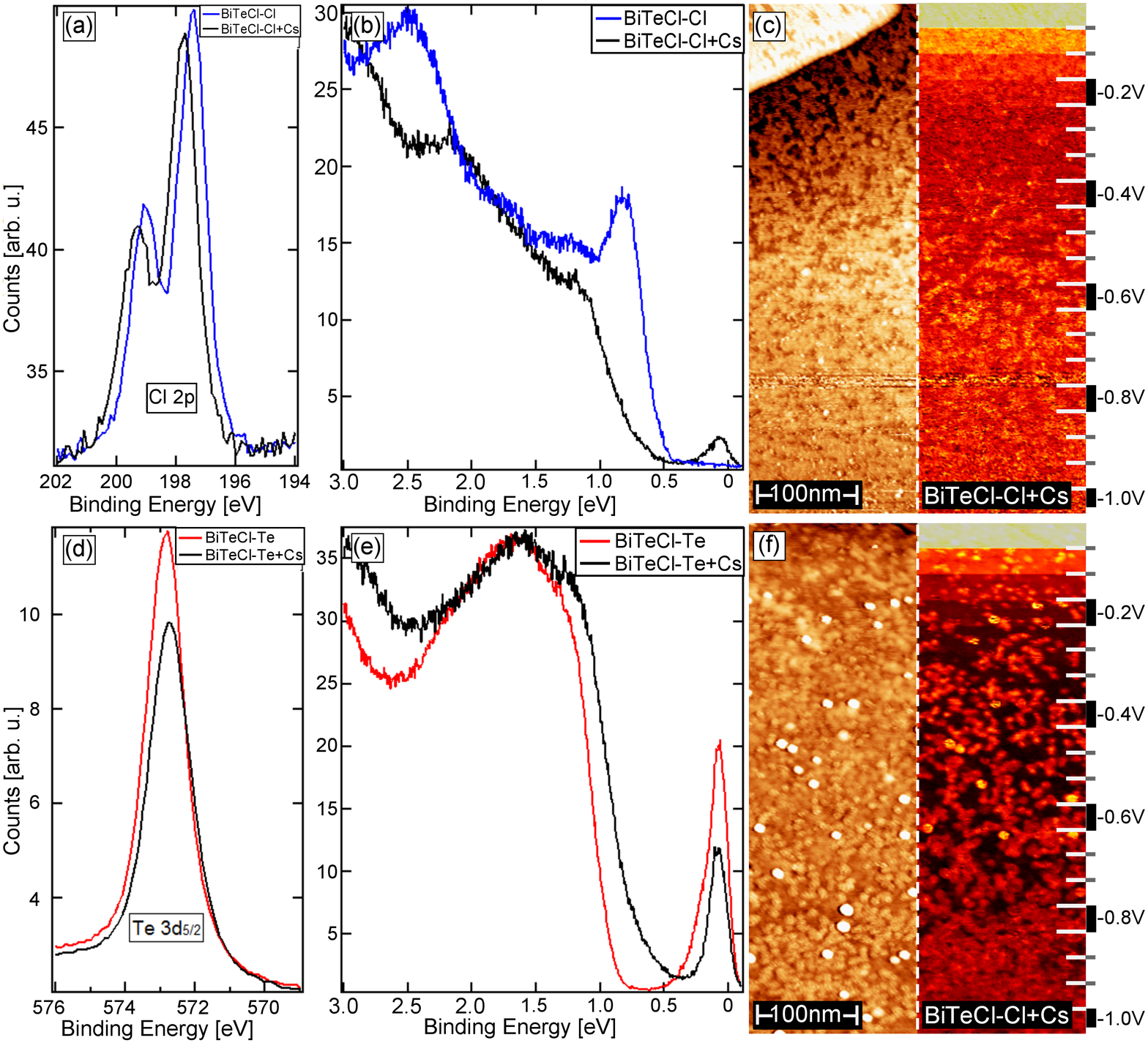}
\end{centering}
\caption{Effect of Cs-adsorption on BiTeCl(0001). (a) and (d) show core level spectra measured before and after Cs deposition on a Te- and a Cl-terminated surface, respectively. Corresponding valence band spectra taken at the $\bar\Gamma$-point are shown in (b) and (e) ($h\nu =$~21.2 eV). STM images and d\textit{I}/d\textit{V} maps acquired after deposition of Cs on Te- and Cl-terminated BiTeCl are given in (c) and (f).}
\label{Figure4}
\end{figure*}
%==================================================================

Fig.~\ref{Figure3} shows ARPES data obtained for BiTeBr and BiTeCl(0001) surfaces. The band structures vary greatly between Te- and Br/Cl-terminated surfaces, but, for a given termination, are similar for both materials. This is in agreement with previous results \cite{Moreschini}. On the Te-terminated surface we observe a Rashba-split band close to the Fermi level that derives from the conduction band bottom and the onset of valence band states at a binding energy of approximately 1\,eV. We note that only one set of parabolic bands is visible in our data whereas previous studies observed two to three sets of bands \cite{Sakano_PRL,Crepaldi_PRB}. In Refs.~\cite{Sakano_PRL,Crepaldi_PRB} the lowest detected bands have their minima below -0.4\,eV while in our case at roughly -0.2\,eV. This could point to a different n-type doping at the surface or in the bulk. Another possible explanation are strong cross section effects with excitation energy which were reported recently \cite{Crepaldi_PRB}. For the Br/Cl-terminated surface conduction band states do not appear at the Fermi level due to p-type band bending as well as no surface states emerge near the valence band in agreement with earlier ARPES measurements on BiTeCl and in contradiction with a theoretical prediction \cite{Landolt_NJP}. The onset of spectral weight derived from the valence band lies at binding energies of approximately 0.7\,-\,0.8\,eV.

The electronic structure determined by ARPES is in fair agreement with the d\textit{I}/d\textit{V} maps in Fig.~\ref{Figure1}, concerning, e.g., the presence or absence of surface states at the Fermi level depending on termination.
In accordance with previous findings for BiTeI we observe significant time-dependent shifts to higher binding energies in the electronic structure of the $X$-terminated surfaces while those are much reduced for the Te-termination \cite{Fiedler}. This can be attributed to residual gas absorption that is enhanced for the $X$-terminations, as already suggested by our STM data. More rapid energy shifts were observed during operation of the He lamp, possibly as a result of hydrogen adsorption, which might explain the discrepancy between the valence band offsets determined by ARPES (Fig.~\ref{Figure3}) and by the d\textit{I}/d\textit{V} maps in Fig.~\ref{Figure1} as well as the absence of the surface states on the $X$ terminations.

Similar to the XPS spectra in Fig.~\ref{Figure2} also the ARPES data in Fig.~\ref{Figure3} reflect the complete surface area of our samples because the spot sizes of the light sources exceed the lateral sample dimensions. The results therefore confirm the single termination of BiTeBr and BiTeCl on a macroscopic scale, in line with the STM data in Fig.~\ref{Figure1}. This excludes any considerable appearance of different crystal phases. The measured band structures show no topological surface state that would bridge the gap between valence and conduction bands, excluding a possible topological insulator phase in BiTeCl \cite{Chen,Yan}.

%==================================================================
\begin{figure*}
\includegraphics[width=0.75\textwidth]{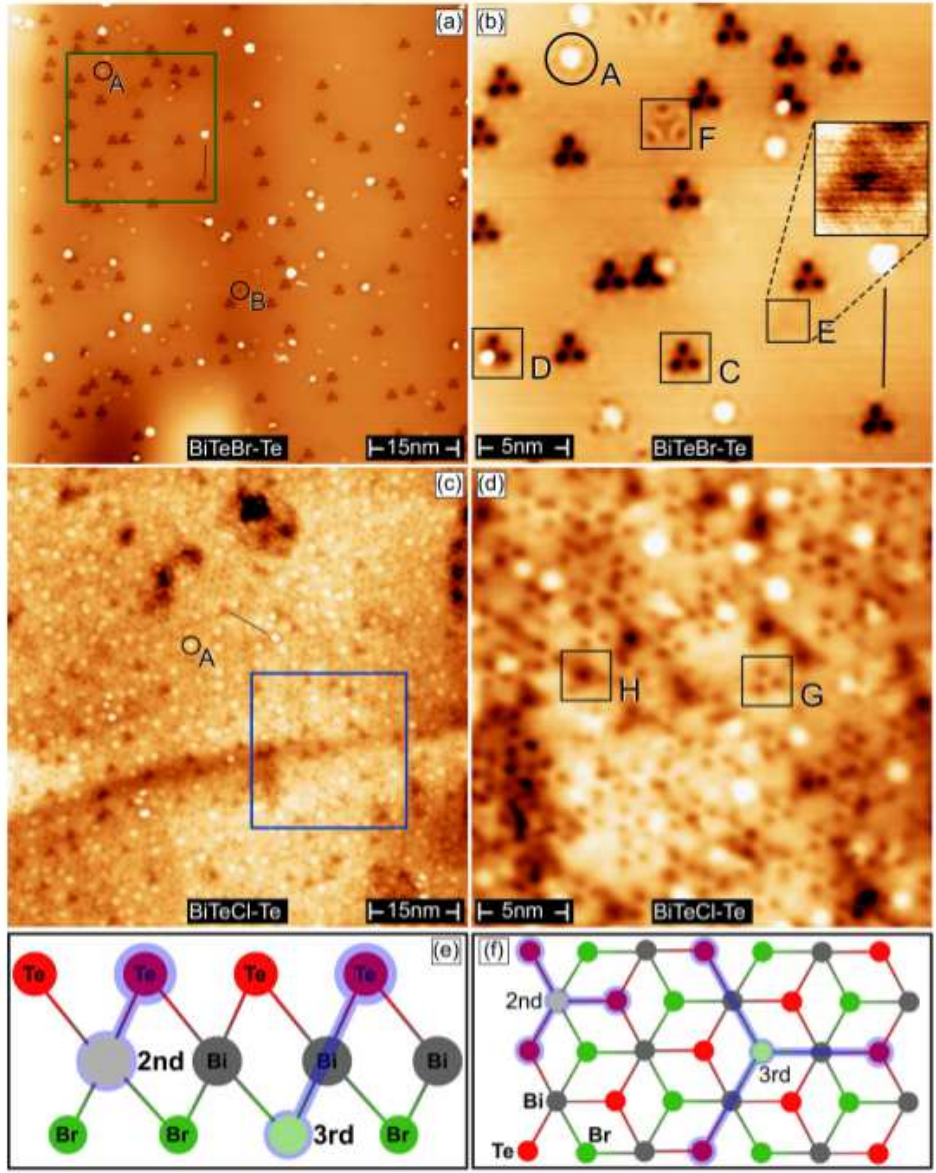}
\caption{LT-STM measurements, all scans are performed at $T$\,=\,5\,K, V\,=\,1\,V and I\,=\,10\,pA. (a) The Te termination of BiTeBr shows the lowest defect density of the BiTe$X$ family. (b) is the zoom in of (a) at the green square, we can find mainly one type of surface defects, three different types of third layer defects with an additional variation, but no second layer defects could be found. (c) shows the Te termination of BiTeCl, the defect density is the highest of the BiTe$X$ compounds. (d) shows the zoom in of (c), one can find at least two different types of defects, others might be covered. (e) side view of a hard ball sketch of BiTeBr-Te. Second (2nd) and third (3rd) layer defects and their effect on nearest neighbor atoms are indicated schematically. (f) top view of a hard ball sketch of BiTeBr-Te. 2nd layer defects would result in three neighboring Te atoms with different contrast while 3rd layer defects mainly affect three next-nearest neighbor Te atoms, as can be seen in defect C in Fig.~\ref{Figure5}(b).}
\label{Figure5}
\end{figure*}
%==================================================================

Since the electronic structure of BiTe$X$ near the surface is highly
termination-dependent it is of interest to investigate additional
possibilities to modify the surface electronic properties.
Fig.~\ref{Figure4} summarizes the influence of Cs adsorption on
the surfaces of BiTeCl. Surprisingly, we observe energy shifts in
the spectra into opposite directions for the two terminations: While
for the Cl-terminated surface the features shift to higher binding
energy - as expected for adsorption of alkali species
\cite{Crepaldi_PRL, Fiedler, Seibel} - they shift to lower
binding energy for the Te-terminated surface. This trend is observed
in the valence band [see Fig.~\ref{Figure4}(b),(e)] and in the
core levels [see Fig.~\ref{Figure4}(a),(d)]. The positive energy
shift on the Te-terminated surface is rather unusual and may occur
in the present case due to clustering of the Cs adsorbates, as
observed by STM in Fig.~\ref{Figure4}(f). In Ref.\,\cite{Fiedler}
we showed for BiTeI that the diffusion length of Cs atoms at
room-temperature is considerably higher for Te- than for
I-terminated surfaces \cite{Fiedler}, which could explain the strong
clustering observed in Fig.~\ref{Figure4}(f). For the
Cl-termination the appearance of the Cs-induced structures in STM is
different and reveals flatter areas with reduced d\textit{I}/d\textit{V}
signal (see Fig.~\ref{Figure4}(c)). As seen in
Fig.~\ref{Figure4}(b) the conduction band minimum shows up below
the Fermi level upon Cs deposition on the Cl-terminated surface,
indicating that it is located slightly above the Fermi level for the
pristine surface. In summary, the results indicate that the surface
termination can considerably affect the adsorption behavior of
adatoms and the resulting influence on the electronic structure,
which might be of relevance, e.g., for interfacing BiTeX with other
materials. Similar effects as presented here for Cs/BiTeCl were also
observed for Cs/BiTeBr (not shown), namely an energy shift to higher
binding energies on the Br-termination and a clustering of Cs
on the Te-termination in combination with an energy shift to lower binding energies.

\subsection{Atomic defects}

After identifying the surface termination, we re-glued the samples with a top-post
and moved them to a separate LT-STM, operated at $T$\,=\,5\,K, to cleave them again.
Fig.~\ref{Figure5} shows data obtained at a positive gap voltage,
usually resulting in increased (decreased) contrast for defects that
act as electron donors (acceptors)
\cite{Jiang}.

If we assume that the sample only consists out of three elements,
for example Bi, Te and Br, three kinds of defects may appear, e.g.
in the Br layer: a vacancy, a Te antisite and a Bi antisite. We
expect that the electronegativity behaves as
Bi\,\textless\,Te\,\textless\,Br (as shown in our DFT calculations)
and that charge of two neighboring atoms is transferred from the one
with lower to the one with higher electronegativity. The atomic
radii behave as Bi\,\textgreater\,Te\,\textgreater\,Br.
One can assume that it is more likely for a vacancy to
be substituted by a smaller atom than forming an antisite with a
larger atom.

In another publication we showed a 400\,nm$^{2}$ scan of the Te termination of
BiTeI \cite{Fiedler} which revealed defect densities of roughly 7.5/(100\,nm$^{2}$) in the third layer
(I) and 2.5/(100\,nm$^{2}$) in the first layer (Te).
Fig.~\ref{Figure5}(a) shows the Te termination of BiTeBr (scan
area 75\,nm$\times$75\,nm) measured at 1\,V gap voltage and
10\,pa tunneling current. With the same method \cite{Jiang}, we can
identify defect densities of about 2.5/(100\,nm$^{2}$) in the third layer
(Br) and 1.3/(100\,nm$^{2}$) in the first layer (Te). No defects
in the second layer (Bi) have been found.

Adsorbates, marked by a black arrow, appear to be around 2.5\,nm high and vary in shape,
while defects labeled (A) are only 25\,pm high and 1\,nm in diameter.
They show an increased contrast and in the zoom-in in Fig.~\ref{Figure5}(b) one further recognizes that the atoms around the
defect center appear darker. This is an indication for a local charge
transfer from the surrounding to the defect atom.
Defect (B) shows a reduced DOS indicating a charge transfer from the defect to the surrounding.
Comparing the defects (A) and (B) by means of total numbers and relative contrast,
we conclude that (A) is a Br antisite while (B) is most likely a Bi antisite or a vacancy.

Now we analyze the three different third layer
defects by means of total number and relative contrast. Defect (C)
appears most often and features the highest contrast. Since the
third layer of the Te termination of BiTeBr is Br, having the
smallest atomic radius and largest electronegativity, a Br vacancy
could be a reasonable candidate.
Furthermore, the basic structure of defect (D) is the same defect as (C) with an additional
atom on top. A possible explanation is a Br atom which remains on
the surface after the
cleaving process. Defect (E) appears less often than (C) but more often than defect
(F) and has the lowest contrast. The atomic radius of Br is closer
to Te than to Bi, which would lead to a Te antisite in the Br layer.
Also the fact that the contrast is weak could be due to the smaller
difference in electronegativity of Te and Br compared to Bi. (F) is
the defect that appears most rarely, which may indicate a Bi
antisite in the Br layer. The high contrast contradicts this
assumptions, but a closer comparison between (C) and (F) shows an
inversion of the contrast. While the center of defect (C)
shows a higher DOS than the direct surrounding, for (F) the situation is opposite: 
a low intensity in the center with a bright surrounding. If we
expect a Bi antisite in the Br layer, the Bi would donate an
electron, which would result in a higher DOS at the location of the
defect \cite{Jiang}. Also the center of defect (E) shows a
dark contrast with a brighter surrounding which would be in line
with our assumptions, since both Bi and Te are less electronegative
compared to Br, so they would act as electron donors.

Fig.~\ref{Figure4}(e) and (f) provide side- and top-view sketches of particular atomic defects in the second a third atomic layer, respectively. While a defect in a certain
layer affects nearest neighbors (NN) atoms, the resulting pattern on the surface gets 
more extended the deeper the defect is located. A second layer defect (2nd) would result in
a contrast change of three NN atoms on the surface.
A third layer defect (3nd) results in a contrast change of three next-nearest neighbor surface atoms,
as can be seen in Fig.~\ref{Figure4}(b) defect C. Defects like E and F appear, when the third layer defect (Br)
influences the NN (Bi / 2nd layer) differently, e.g. acting as an electron donor instead of an electron acceptor.
The result is a Bi atom acting like a 2nd layer defect and therefore
in three Bi atoms influencing three neigboring atoms (Te) each.

Like on BiTeI \cite{Fiedler} no defects below the third layer could be found, possibly due to the van-der-Waals gap.
The whole surface seems to be corrugated, as can be seen on the
bottom part of Fig.~\ref{Figure5}(a) at the dark and bright area, which might be the result
of screw dislocations. If we compare the Te termination of BiTeBr and
BiTeI, the defects E and F of Fig.~\ref{Figure5}(a) are very similar to the defects E and F
from Fig.\,2 in Ref.\,\cite{Fiedler}, which could also be Te and Bi antisites.

The defect density in BiTeCl
[Fig.~\ref{Figure5}(c)] is much higher as compared to BiTeBr. It is
difficult to find a vacancy in the first layer but adsorbates (black
arrow) and antisites (A) can frequently been found. Fig.~\ref{Figure5}(d) is the
magnified view of the blue-framed square
shown in Fig.~\ref{Figure5}(c). It is hard to
point out certain defects but (G) and (H) probably represent different third layer
defects, most likely a vacancy and a Te antisite, respectively.

So far measurements in the LT-STM were only successful for the Te-terminated surfaces of BiTeBr and BiTeCl. 
However, third layer defects of Te should be equal to first layer defects of $X$, as long as they are not induced by the cleaving process. This would mean at least for BiTeBr that the Bi layer is almost free of defects and that the Te-layer has less defects than the Br layer.

\section{Discussion}

Comparing the three BiTe$X$ compounds the most obvious difference is the presence ($X$ = I) or absence
($X$ = Cl, Br) of stacking faults in the bulk crystal structure
resulting in surfaces with mixed or single terminations,
respectively. On the atomic scale, however, BiTeCl stands out with a
considerably larger defect density than the two other compounds.
Hence, in this respect BiTeBr currently appears to be the material
with the most homogeneous structural properties. This finding nicely
complements comparative studies of the surface
electronic properties of BiTe$X$ that suggests BiTeBr as the
best candidate for possible future applications
\cite{Eremeev_NJP,Moreschini}.

We further note that a possible
migration of Bi atoms into the topmost Te-layer was speculated to
occur in all three BiTe$X$ compounds based on the observation of a
second component in the Bi $5d$ core level signal for Te-terminated
surfaces \cite{Moreschini}. In our STM measurements for BiTeBr,
however, such defects involving the first (Te) and the second (Bi)
layer are not found. On the other hand, also no additional component
in the Bi core level spectra is observed in the present study, in
agreement with a previous report on BiTeCl \cite{Landolt_NJP}.

The role of structural defects is furthermore important for a basic understanding of the electronic properties in BiTe$X$. For BiTeCl a lift-off during the cleaving process of a thin free-standing layer (around 1 unit cell) that remains loosely on the crystal surface has been proposed to give rise to the Rashba-split surface bands observed in ARPES and to mask the presence of a topological state on the intrinsic surface \cite{Chen}. This scenario is not supported by the present combined STM and ARPES results that show step edge heights of the surface terraces matching the bulk unit cell and, at the same time, provide no indication of topological surface bands. It is furthermore worth noting that, while the atomic defect density observed here in STM is considerably higher for BiTeCl than for BiTeBr, the quality of the ARPES data turns out to be comparable and also the measured band structures are very similar. This observation is in contrast to a recent investigation of BiTeCl that concluded qualitative changes in the electronic structure depending on the amount of defects near the surface \cite{Yan}.

The broken inversion symmetry in BiTe$X$ in combination with the
high electronegativity of the halogen atoms is assumed to induce a
net dipole moment in the bulk unit cell \cite{Chen,Kohsaka} that, in
turn, gives rise to n- or p-type band bending at the surface
depending on termination \cite{Crepaldi_PRL}. The proposed
microscopic picture of the charge distribution is often based on a
covalently bound (BiTe)$^{+}$ bilayer that couples ionically to the
adjacent $X^{-}$ layer
\cite{Shevelkov,Crepaldi_PRL,Moreschini,Butler}. However, the
bonding character has also been viewed as ionic for, both, Bi-Te and
Bi-$X$ based on the fact that the valence (conduction) band is to
most extent Te/$X$ (Bi) derived which indicates significant charge
transfer from Bi to Te and $X$ \cite{Zhu}. In some calculations even
a larger charge transfer to Te than to $X$ has been obtained
\cite{Chen, Ma}. Direct experimental information
on this issue has so far been scarce.  The present XPS measurements
indeed point to a substantial charge donation from Bi to Te and $X$
which is in line with our first-principles calculations of the local
atomic charges. On the other hand, the large work function
differences between Te- and $X$-terminated surfaces confirm the
presence of a dipole moment in the unit cell and, thus, support the
view of a (BiTe)$^{+}$ block with positive net charge that forms a
polar bond with the $X^{-}$ layer.

\section{Summary}
We have presented a comparative study of the structural and
electronic surface properties of the non-centrosymmetric
giant-Rashba semiconductors BiTe$X$(0001) ($X$ =  Cl, Br, I).
Cleaving of single-crystalline samples exposes macroscopically
homogeneous surfaces with Te- and $X$-termination for BiTeCl and
BiTeBr, in contrast to BiTeI where bulk stacking faults are known to
give rise to mixed surface terminations. STM and XPS data confirm
the unit cell heights and atomic stacking orders that are expected
from the bulk crystal structure. The electronic band structures
measured by ARPES differ considerably depending on surface
termination, but in no case topological surface states are observed.
The chemical bonding in BiTe$X$ is found to be characterized by
substantial charge transfer from Bi to Te and $X$.
However, based on work function measurements we also obtain evidence for ionic bonding
between (BiTe)$^{+}$ bilayers and $X^{-}$ layers, whereas the polarity of the bond
increases with rising electronegativity of the halogen atom.

\section{Acknowledgements}
This work was financially supported by the
Deutsche Forschungsgemeinschaft through FOR1162 and partly by the Ministry of
Education and Science of Russian Federation (Grant No. 2.8575.2013),
the Russian Foundation for Basic Research (Grants No. 15-02-01797, 15-02-02717)
and Saint Petersburg State University (project 11.50.202.2015).

\end{document}